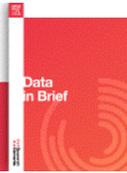

# ARTICLE INFORMATION

**Article title**

Dataset of turbulent flow over interacting barchan dunes

**Authors**

Jimmy G. Alvarez

Danilo S. Borges

Erick M. Franklin*

**Affiliations**

Faculdade de Engenharia Mecânica, Universidade Estadual de Campinas (UNICAMP), Rua Mendeleyev, 200, 13083-860, Campinas-SP, Brazil

**Corresponding author's email address and Twitter handle**

erick.franklin@unicamp.br

**Abstract**

Barchans are dunes commonly found in dune fields on Earth, Mars and other celestial bodies, where they can interact with each other. This article concerns experimental data for the flow over subaqueous barchans that are either isolated or interacting with each other. The experiments were carried out in a transparent channel of rectangular cross section in which turbulent water flows were imposed over either one single or a pair of barchans. The instantaneous flow fields were measured by using a low-frequency PIV (particle image velocimetry) and high-frequency PTV (particle tracking velocimetry). From the PIV and PTV data, the mean flow, trajectories, and second-order moments were computed, which are included in the datasets described in this paper, together with raw data (images), instantaneous fields, and scripts to process them. The datasets can be reused for benchmarking or for processing new images generated by other research groups.

# SPECIFICATIONS TABLE

| Subject | Earth and Planetary Sciences / Earth-Surface Processes |
|---|---|
| Specific subject area | Disturbed flow over dunes interacting with each other. |

| **Data format** | Raw PTV images, processed PIV vectors (using the software DaVis), processed data in MatLab format (matrices of flow fields and vectors of flow profiles), and MatLab scripts for both processing the raw data and post-processing processed data. |
| --- | --- |
| | Raw, Analyzed, Filtered |
| **Type of data** | .vc7 (PIV vectors in DaVis –LaVision format) |
| | .tiff (raw PTV images) |
| | .mat (processed data in MatLab format) |
| | .m (MatLab scripts) |
| | Image, Movie, Matrices, Numerical Code |
| **Data collection** | The PIV data were collected using a charge-coupled device (CCD) camera, and the PTV data with a camera of complementary metal-oxide semiconductor (CMOS) type, both placed perpendicular to a laser plane. For the PIV, we used a dual-cavity Nd:YAG Q-switched laser emitting pulses at 2×130 mJ, and for the PTV we used a continuous-0.3-W laser. The CCD camera had a low (4 Hz) time resolution, while the CMOS camera had a high time resolution (400 Hz). Maximum resolution is space was either 2048 px × 2048 px or 2560 px x 1600 px, respectively. |
| **Data source location** | The data were collected at the School of Mechanical Engineering of the University of Campinas (UNICAMP), Campinas, São Paulo, Brazil. |
| **Data accessibility** | Repository name: Mendeley data repository |
| | Data identification numbers: 10.17632/hyrfnpf8hc.1, 10.17632/jn8dddnj26.1, and 10.17632/r92y33x5pp.1 |
| | Direct URLs to data: |
| | https://data.mendeley.com/datasets/hyrfnpf8hc/1 |
| | https://data.mendeley.com/datasets/jn8dddnj26/1 |
| | https://data.mendeley.com/datasets/r92y33x5pp/1 |

# VALUE OF THE DATA

- The data is useful for understanding the dynamics of dunes found in nature.
- Researchers, geophysicists, and civil entrepreneurs may benefit from these datasets.
- The data can be reused to predict the behavior of barchan fields and assess the impact onto human activities on Earth and other celestial bodies.

# BACKGROUND

Barchans are dunes commonly found in dune fields on Earth, Mars and other celestial bodies, where they can interact with each other. Although they share roughly the same morphology and dynamics, the scales differ depending on the environment they are in: one kilometer and millenniums for Martian dunes, hundreds of meters and years for eolian dunes, and centimeters and minutes for subaqueous dunes [1].

One important mechanism for barchan-barchan interactions is the fluid flow, which, after being disturbed by the upstream barchan, impacts the downstream one [2-4]. In this paper, we present experimental datasets concerning the turbulent flow over barchan dunes. Because of the much smaller and faster scales of subaqueous barchans, we carried out experiments in a water channel. For that, we made use of particle image velocimetry (PIV) and particle tracking velocimetry (PTV), as described next. The generated datasets can be useful for researchers willing to understand how flow disturbances affect barchans within a dune field.

# DATA DESCRIPTION

The datasets of this work can be found in Refs. [5-7], being available on Mendeley Data. They are divided in three different addresses : https://data.mendeley.com/datasets/hyrfnpf8hc/1, https://data.mendeley.com/datasets/jn8dddnj26/1, and https://data.mendeley.com/datasets/r92y33x5pp/1 for, respectively, PIV experiments over single barchans, PIV experiments over the downstream barchan (in the case of interacting barchans), and PTV experiments over the single and downstream barchans.

The datasets [5] and [6] (for PIV) have the same structure. Folders "Single Dune_PIV_vc7"and "Dunes_in_Interaction_PIV_vc7" contain vector fields of the instantaneous flow in vc7 format (a file extension native to the LaVision software DaVis). Each field was computed by cross correlating two PIV images using the software DaVis, and each folder also includes an image (.jpg) as an example of a visualization captured during the PIV tests. The vc7 files are stored in three subfolders named "Part_1" to "Part_3", each one accounting for one of the different regions imaged (described next in section EXPERIMENTAL DESIGN, MATERIALS AND METHODS). Due to size limitations, only 200 (10% of the 2000 vector fields) are stored in each of these subfolders. The folder "Matlab codes" contains Matlab scripts for post-processing the vc7 files in order to obtain the mean flow and second-order moments, and the folder "Data_MatLab_format" contains computed fields and profiles in MatLab format (.mat). Finally, the folder "Movies" contains movies showing the setup and how the tests were conducted (filmed with an independent camera as one experiment was going on). Detailed instructions on how to use the files are listed in the datasets.

The dataset [7] (for PTV) contains 6 folders. Folders "Images_Single_Dune_400 Hz" and "Images_Dunes_in_interaction_400 Hz" contain, respectively, raw images for the single and downstream barchans, folder "MatLab Codes" contains Matlab scripts for detecting and tracking particles, folders "Post-processing_Single" and "Post-processing_Interaction" contain MatLab scripts for post-processing, respectively, the flow fields over either the single or downstream barchans, and the folder "Extras" contains a movie showing the setup and how the tests were conducted (filmed with an independent camera as one experiment was going on). Detailed instructions on how to use the files are listed in the dataset.

# EXPERIMENTAL DESIGN, MATERIALS AND METHODS

The experimental setup consisted basically of a water reservoir, two centrifugal pumps, a flowmeter, a flow straightener, a 5-m-long closed-conduit channel, a settling tank, and a return line. With the pumps turned on, the water flowed in turbulent regime, following the order just described. The channel was transparent, had a 160-mm-wide and 50-mm-high rectangular cross section, and its last 2 m consisted of the test and discharge sections (1-m-long each). The experimental setup is similar to those described in Refs. [3,4, 8,9]. Figure 1a shows a layout of the test section.

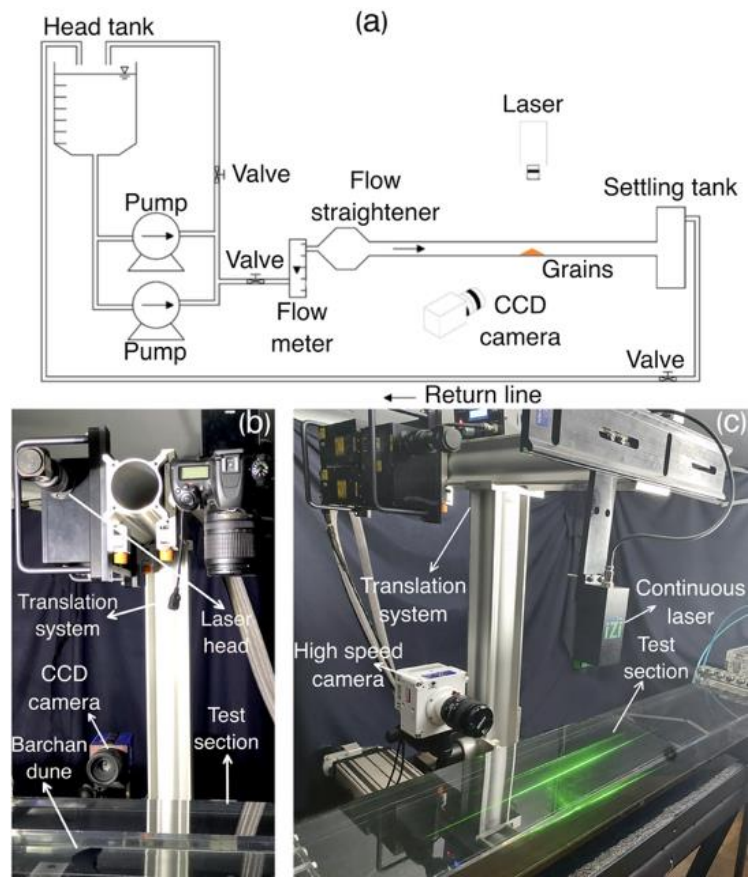

Figure 1: (a) Layout of the experimental device. (b)-(c) Photograph of the test section for (b) PIV and (c) PTV tests.

Prior to each experiment, controlled grains (glass spheres with diameter within 0.15mm ≤ d ≤ 0.25 mm) were poured in the channel (already filled with water), forming one conical pile. Afterward, the pile was deformed into a barchan dune by imposing a water flow (by turning on the pumps). Once a barchan dune was developed, the flow was reduced below the threshold for grains motion (stopping the barchan but not the flow) and the PIV or PTV experiments over the single dune were carried out. With all PIV or PTV experiments conducted, the flow was stopped, the channel access opened, and more grains were poured inside forming a second pile that was deformed into a second barchan (upstream the first one) once the water flow was imposed. When both barchans were in the desired configuration, the flow was reduced again below the threshold for grains motion, and PIV or PTV experiments were carried out for the flow over the downstream barchan. Figure 2 shows top views of barchans in both configurations (single and interacting barchans), where we indicate by green lines the positions where the laser plane impacted the barchan surface. The glass spheres forming the barchans were black in order to reduce undesirable reflections.

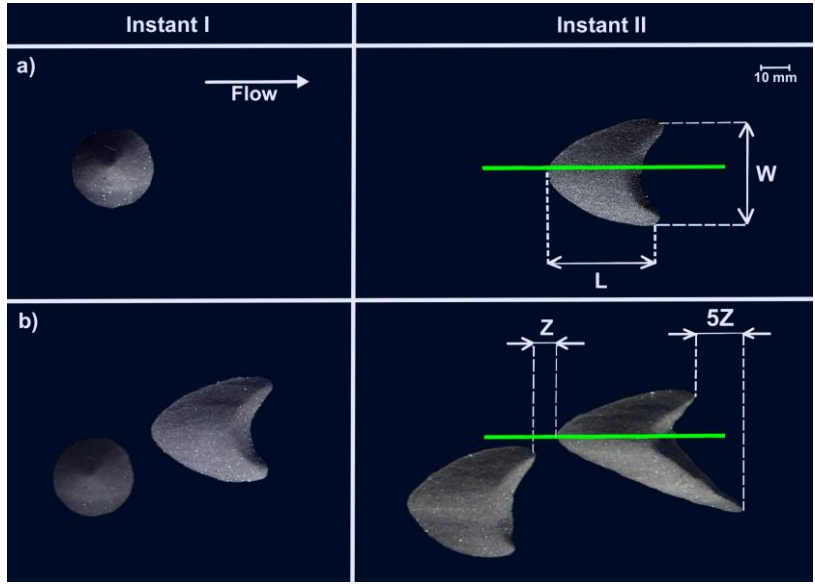

Figure 2: Top-view images of dunes used in the tests. The green line represents the impact line of the laser sheet.

In the experiments, the laser plane was in the vertical position, hitting the barchan of interest along its respective centerline. For the PIV we used a dual-cavity Nd:YAG Q-switched laser capable of emitting pulses at 2×130 mJ at a maximum frequency between pairs of 15 Hz, and for the PTV we used a continuous-0.3-W laser. The lasers were mounted on a traveling system, on which we also mounted a camera with field of view perpendicular to the laser plane. For the PIV we used charge-coupled device (CCD) camera with resolution of 2048 px × 2048 px and maximum frequency of 10 Hz, while for the PTV we used a CMOS camera capable of acquiring images at 800 Hz at its maximum resolution, which was 2560 px x 1600 px. Figures 1b and 1c show the test section with cameras and lasers for the PIV and PTV experiments, respectively, for which we set 4 Hz for the PIV and 400 Hz for the PTV experiments. Short movies showing the setup and how the tests were conducted are available in the folder "movies" of datasets [5] and [6], and in the folder "extras" of dataset [7].

The field of view of PIV tests was of 52.5 mm x 52.5 mm for a region of interest (ROI) of 2030 px x 2030 px, and images were processed with the LaVision software DaVis. We used interrogation zones of 16 px × 16 px and 50% of overlap, corresponding to a spatial resolution of 0.21 mm x 0.21 mm. Examples of PIV vectors generated by DaVis for the single and downstream barchans are stored in folders "Single Dune_PIV_vc7" and "Dunes_in_Interaction_PIV_vc7", in their respective datasets ([5] and [6], respectively). Because the field of view of PIV was smaller than the size of barchans, three different regions were imaged in order to cover the whole centerline of barchans. The regions are numbered 1 to 3 going from upstream to downstream, corresponding the subfolders "Part_1" to "Part_3" inside folders "Single Dune_PIV_vc7" and "Dunes_in_Interaction_PIV_vc7". Matlab scripts for post-processing the vc7 files in order to obtain the mean flow and second-order moments are in folder "Matlab codes", and the computed fields and profiles in MatLab format (.mat) are in folder "Data_MatLab_format", in both datasets.

The field of view of PTV experiments was of 99 mm x 54 mm for a region of interest (ROI) of 2547 px x 1251 px (so that 1 px corresponded to approximately 0.04 mm), and the images were processed with MatLab scripts written in the course of this work. For tracking the seeding particles, the code first identified the particles in each image, and then looked for the minimum displacements between them along all the images, using a Kalman Filter [10] to account for disappearing/reappearing particles. This is illustrated in Fig. 3. The PTV data is available on dataset [7], in which raw images for the single and downstream barchans are stored in folders named "Images_Single_Dune_400 Hz" and "Images_Dunes_in_interaction_400 Hz", respectively, while Matlab scripts for detecting and tracking particles are available in the folder "MatLab Codes". Finally, MatLab scripts for post-processing the flow fields over either the single or downstream barchans are available in folders "Post-processing_Single" and "Post-processing_Interaction", respectively.

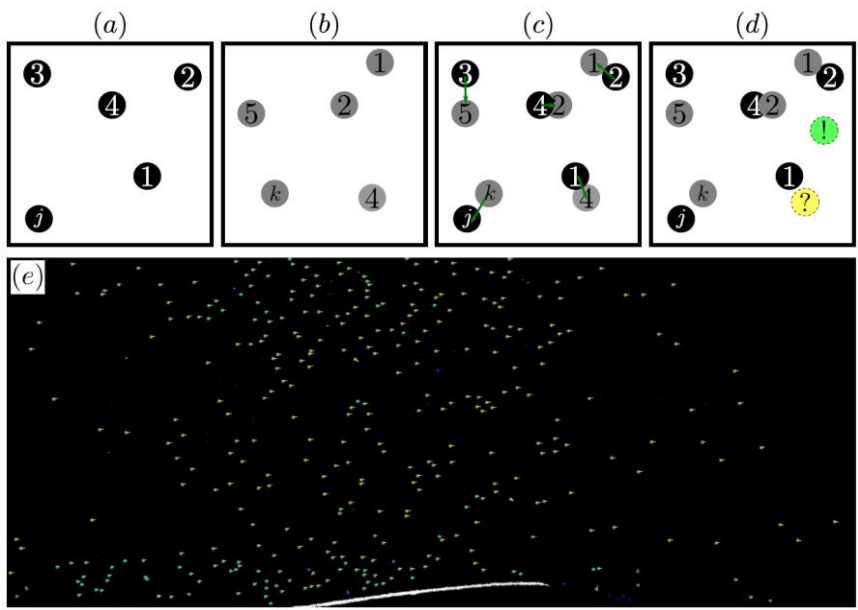

Figure 3: (a)-(b) Identification and labelling of particles in consecutive frames, (c) tracking pf particles (minimum displacement) when there is no appearance/disappearance problem, and (d) tracking of particles when one particle disappears (yellow) and another one appears (green). (e) Example of motion of detected particles in two consecutive images superposed with the first image.

# LIMITATIONS

Not applicable.

# ETHICS STATEMENT

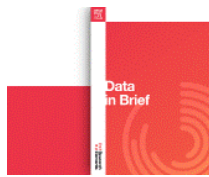

# CRediT AUTHOR STATEMENT

**Jimmy Gabriel Alvarez:** Conceptualization, Methodology, Software, Validation, Formal Analysis, Investigation, Data Curation, Writing - Review & Editing, Visualization. **Danilo S. Borges:** Methodology, Software, Formal Analysis, Investigation, Writing - Review & Editing, Visualization. **Erick M. Franklin:** Conceptualization, Methodology, Formal Analysis, Investigation, Resources, Writing - Original Draft, Writing - Review & Editing, Supervision, Project Administration, Funding acquisition.

# ACKNOWLEDGEMENTS

Funding: This work was supported by São Paulo Research Foundation (FAPESP) [grant numbers 2018/14981-7, 2020/15624-3, and 2022/01758-3], and for CNPq [grant number 405512/2022-8]. The authors thank Estaban Cúñez for his assistance with some of the experiments and image editing.

# DECLARATION OF COMPETING INTERESTS

The authors declare that they have no known competing financial interests or personal relationships that could have appeared to influence the work reported in this paper.

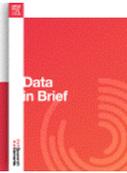